# The Electrical-Thermal Switching in Carbon Black-Polymer Composites as a Local Effect

D. Azulay, M. Eylon, O. Eshkenazi, D. Toker, M. Balberg, N. Shimoni, O. Millo, and I. Balberg

The Racah Institute of Physics, The Hebrew University, Jerusalem 91904, Israel

Following the lack of microscopic information about the intriguing well-known electrical-thermal switching mechanism in Carbon Black-Polymer composites, we applied atomic force microscopy in order to reveal the local nature of the process and correlated it with the characteristics of the widely used commercial switches. We conclude that the switching events take place in critical interparticle tunneling junctions that carry most of the current. The macroscopic switched state is then a result of a dynamic-stationary state of fast switching and slow reconnection of the corresponding junctions.

72.80.Tm, 68.37.Ps, 81.05.Qk

The electrical properties of Carbon Black-Polymer (CBP) composites have been studied intensively for the last sixty years.[1,2,3] A conspicuous reason for this interest, in addition to fundamental physical problems associated with them,[4,5] is the widely used application of these materials for current-limiting resetable switches.[6,7,8] While the operation of these switches involves commercial interests and intriguing physical problems, there were very few open literature publications that discussed the corresponding switching mechanism. For example, one can hardly find reports on the current-voltage characteristics of such devices.[8] It was, however, recently, for the closely related system of Carbon beads-Polymer composites, that such characteristics were reported.[9] In few other systems of metal particle-polymer composites the related phenomenon of breakdown, and in particular its time dependence, have been studied.[10,11] The common understanding of all these observations and the device operation is based on the coupling of Joule heating and positive thermal coefficient of the resistance (PTCR) in these composites. This originates from the considerable expansion of the polymer when the temperature increases towards the melting point.[6] In contrast to the case of the switching phenomenon, there are numerous papers on the PTCR effect, i.e. the orders of magnitude increase of the resistance of the CBP composite upon the (uniform) increase of the temperature.[6,12] The approaches that try to account for this effect vary between "pure" percolation models (i.e., attributing the resistance increase simply to the effective lowering of the conducting-phase content [8,13]) to the increase of interparticle distance (bringing about to a lower tunneling probability[14] or broken contacts[9,10,11]) or to mixtures of the two approaches.[12,15]

These phenomena, in particular the low-to-high resistance switching effect, were studied thus far using macroscopic techniques. The corresponding data were discussed by continuous-macroscopic models,[8,16] by a collective-global electro-thermal (ET) breakdown[9] or by simulations[17,18] of a network that is made of non-linear ET switches. The macroscopic character of the experimental observations and the global nature of the models do not enable however the evaluation of the local nature of the phenomenon and its relation to the macroscopic results. In fact, even a basic parameter such as the time it takes for the "contact" between two CB particles to fuse-off has not been determined. In this letter we present an investigation of the local aspects of the switching behavior in such composites by employing conducting atomic force microscopy (C-AFM).[19,20] This approach is shown to provide further insight into the mechanism leading to the global ET phenomenon.

The basic manifestation of the macroscopic ET effect is that the application of a voltage beyond some threshold value yields a large increase of the macroscopic resistance.[7,8,9,10,11] This effect is exhibited in Fig. 1 by the current voltage (I-V) characteristics of a "device" made of a (50 vol.% low structure[1,14]) CBP composite, which is one of the composites that were studied in the present work. This "device" (0.5 mm thick with an area of 0.5 cm$^2$) had contacts of laminated metal foils on its opposite surfaces. The behavior shown in Fig. 1 is similar to the one found previously on CB[8] and carbon-beads[9] polymer composites under constant voltage conditions. These results present also the time scale that is involved in the macroscopic process, namely, that the resistance increases with time, following voltage-application, for up to about 30 sec, after which it saturates. This shows that under the conditions used, a macroscopic stationary

"switched" state is reached within 30 sec. The question that arises now is whether we obtain after these 30 sec a stationery fused-resistor network or a dynamic situation of an ever opening and closing of individual resistors in the network. Hence the local nature of the fusing process appears to be essential for the understanding of the global switching phenomenon. C-AFM is an effective tool for this task, providing direct information on the conduction-percolation network in metal-insulator composites.[19]

Our CBP samples were prepared from commercial grade polyethylene filled with either 10-23 vol.% of high structure[1] CB (characterized[14] by a low effective percolation threshold, $p_c$ = 9.3 vol.%) or with 46-63 vol.% of low structure[1] CB ($p_c$ = 40 vol.%). The preparation and the electrical properties of these CB composites were addressed previously.[6,14] Briefly, the high structure CB particles are graphite aggregates (about 500 nm long and 30 nm in diameter). The low structure CB that we used are spherical graphite particles with a diameter distribution that is centered around 200 nm. The samples that were utilized in the C-AFM measurements had an area of about 1 cm$^2$ and a thickness of 0.25 mm. For the measurements, a silver paint strip was pasted on the edge of the sample to serve as the counter-electrode to the conducting AFM tip. As test samples for the local switching behavior, we have used Ni-SiO$_2$ granular metal composites[20] (with a wide range of Ni content) that do not exhibit global ET switching.

In our measurements we applied a commercial (NT-MDT solver) AFM, operated in the constant-force mode, to acquire local current-voltage (I-V) characteristics and to map the current distribution through the samples' surface (i.e. between the tip and the counter electrode, separated by more than 2 mm). In parallel, we mapped the topography and carried out material-sensitive friction (lateral force) and stiffness (force modulation) measurements,[21] yielding contrast between the CB and polymer phases. Typically, areas of 5x5 µm² were scanned using a conductive tip (silicon tip that has a curvature less than 35 nm, coated by a 25 nm thick titanium-nitride films). The fact that the size of the tip is smaller than the individual carbon black particle is of great significance, since it means that the current that reaches the electrode is essentially initiated at a single CB particle on the surface of the composite (see below). Thus, the corresponding current path must be via the contact of this surface-protruding particle to the electrically connected chain of CB particles (constituting a "dead end"[22]). The latter merges the backbone of the conducting network, terminating at the counter electrode. In the first stage of our study we monitored the regions ("current islands") through which there is a measurable current.[19] Since each of these "islands" corresponds to a current path between the tip and the electrode, its appearance or disappearance with the change of bias corresponds to the opening or closing of such a tip-electrode percolation-like path.

In Fig. 2 we present two current images taken on a 46 vol% (low structure) CB composite for the same area of a sample, but under the application of two different tip-electrode voltages (5V and 5.6V). With increasing bias, up to about 5 V, we have observed a monotonous increase of the number of "conducting islands" and the total area covered by them (see below). This is expected as more current paths are opened. What is a priori unexpected for regular composites (and indeed not observed in the Ni-SiO₂ composites) is the decrease of the number of current islands and the integrated current upon a further increase of bias, such as between 5 and 5.6 V (in particular note the encircled areas in Fig. 2). This behavior is expected, however, if along the interparticle conduction path, a critical current connection is switched off due to the local Joule heating that causes a huge increase of the tunneling resistance of an individual inteparticle junction, following the local polymer expansion. In particular, considering the typical CB particle geometry, we know that the corresponding current densities to be as high as 100 A/cm$^2$ per measured nA. Since the tip scans a pixel within a msec, we conclude that such a local switching event takes place within that time.

Noting that AFM is a surface probe technique by its very nature, a question arises as to the possibility that the observed local switching events take place at the tip-sample junction. To resolve this issue we correlated the current maps with the surface-sensitive friction and stiffness measurements,[21] as mentioned above. The good correlation between the current and the friction images can be appreciated by comparing Figs. 3(a) and 3(b). In the friction image, the dark islands correspond to the CB particles protruding from the surface where the tip-surface friction happens to be smaller than the tip-polymer friction (brighter region). A similar correlation was observed between the current and stiffness maps. These correlations indicate that for most of the current islands the tip touches a "bare" CB particle directly and consequently the switching cannot take place between the tip and the surface, but rather in the bulk.

We found an independent macroscopic support for this conclusion from our thermal images of another sample (60x10x0.5 mm$^3$) of low structure CBP composite, positioned between two coplanar (60 mm separated) contacts. These images, that will be presented elsewhere, show that under bias application high temperature regions develop in areas (the "hot spots" [18] or "hot zones"[17]) that are well removed from the coplanar electrodes. This indicates that the main global switching, and therefore the local switching events, take place in the bulk and not necessarily at the contacts, as was suggested[11] for similar systems.

The understanding of the results shown in Fig. 2 in terms of a local ET switching effect that was described above is further supported by our local (on "one island") I-V measurements. In Fig. 4 we present two typical characteristics. In the first, there is an apparent (complete) fusing of the current at a given voltage (a) and in the other, the first switch is followed, upon a further increase of the bias, by a second fusing event (b). This and the disappearance of some "conducting islands" in Fig. 2 confirm conclusively that we do not have a moderate resistance increase of all or many resistors, but rather a full fusing of some local resistors. Hence, the global switching effect in Fig. 1 is a result of the complete fusing of *some* individual interparticle contacts. This new observation suggests then the *deterministic* rather than the random-global switching effect, at least under constant low-voltage conditions. In Fig. 4, case (a) can be attributed to the fusing of an interparticle junction while case (b) indicates that a further increase of the applied voltage can bring about the opening of an alternative-parallel current path that is initiated at the same current node. This can be viewed as a flow process in a Cayley tree (or a Bethe lattice[22]) on which there are junctions that are opened or closed at different values of the applied bias.

The above-described results can be accounted for by the competition between the local opening and fusing of tunneling-contacts. The branched (Cayley tree type) structure of the backbone suggests that there is a relatively wide distribution of connections (tunnel-barriers) such that the increase of bias voltage opens more and more connections. On the other hand, at higher voltages, the fusing yields the elimination of more and more connections. This leads to a non-monotonic bias dependence of the total current integrated over the scanned area, as shown in the inset of Fig. 5. We characterize the dominance of the fusing effect by the threshold voltage at which the total current first decreases (marked by the arrow). This threshold must be associated with "critical" junctions in which nearly the whole current passes through, i.e., in the "dead ends" mentioned above or in junctions of lowest local resistance in the current paths ("singly connected-like bonds"). The tunneling nature of the conduction enables then a closing of such junctions as well as the opening of others.

To further demonstrate that the local tunneling mechanism account well for the induced conduction and fusing effect, we show in Fig. 5 the dependence of the threshold voltage on the CB content for high-structure CBP composite (a similar behavior was observed with low-structure CB). The data shown are the result of an average of scans of three different areas on the sample. We see that the larger the CB content the smaller the threshold voltage. This is in accordance with the simple expectation that the interparticle tunneling barriers become narrower with the increase of the CB content. This has two effects. First, due to the smaller corresponding resistance, the current and heat dissipation in the junctions (for a given bias) is larger in the case of the higher CB content. Second, the narrower junctions require a smaller Joule heating for the fusing due to the smaller amount of the polymer material in them.

In conclusion, we have shown that under low constant-voltage conditions individual tunneling barriers are completely switched-off for tunneling conduction. These barriers appear to be in the current-leading "dead ends" (that start at the surface-protruding CB particles) or in the "singly connected-like" junctions of the percolation backbone. The observed global effect under these conditions is a result of uncorrelated opening and closing of these barriers. Above a certain bias the latter effect dominates and the global fusing action takes place. The corresponding stationary state represents then a dynamic situation of different ensembles of switches at each time. While the individual switching time is shorter than a msec, the formation of the stationary state (that also involves a local on-and-off cooling process) takes tens of seconds. In turn, our results also indicate that the PTCR effect is mainly driven by the change in interparticle resistance rather than by the "geometrical" effective decrease of the CB content.

Acknowledgements:

The authors would like to thank M.B. Heaney and M. Wartenberg for the samples used in this study. This work was supported by the Israel Science Foundation.

Figure 1. Typical current-voltage characteristics of an ET switching device made of a composite of 50 vol.% low structure CB and a polymer. The current is read 1 (open circles), 30 or 300 sec. (full circles) after bias onset.

Figure 2. Two 5x5 mm2 current images taken on the same area of a 46 vol% low structure CBP sample, for two different voltages, as indicated. Note the disappearance of some of the conducting islands (encircled for 5V), and the general current reduction, upon bias increase (the current scale is shown to the right).

Figure 3. 1x1 mm2 current image (a) and frictions image (b) taken simultaneously on a sample of (high structure) 23vol.% CBP composite.

Figure 4. Two typical kinds of I-V characteristics that exhibit a local ET switching. In the first, (a), there is a single switching event and in the other, (b), the first switching is followed by another one. These characteristics were obtained on an 18 vol.% high structure CBP composite.

Figure 5. The dependence of the first threshold voltage on the CB content (for high structure CB-polymer composites). The inset shows the dependence of the total area of conducting islands on bias voltage, for a 15 vol.% CB sample, and the arrow marks the first threshold voltage.

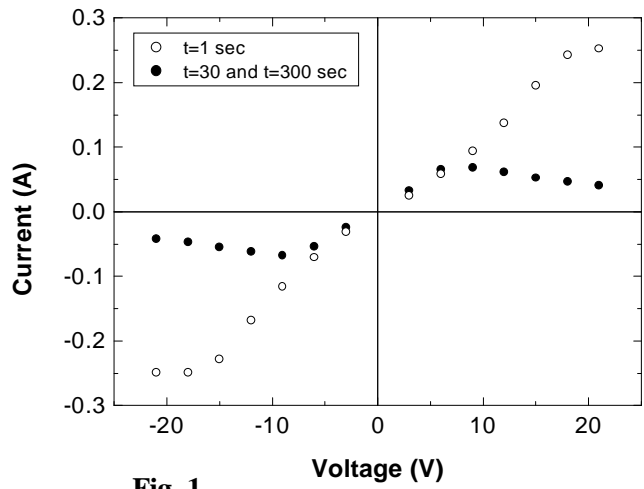

**Fig. 1**

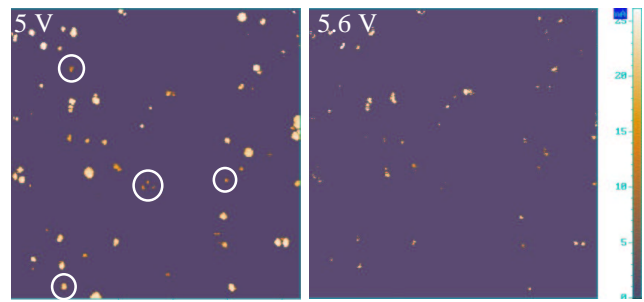

**Fig. 2**

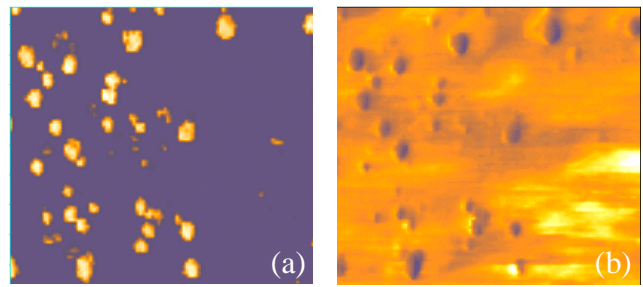

**Fig. 3**

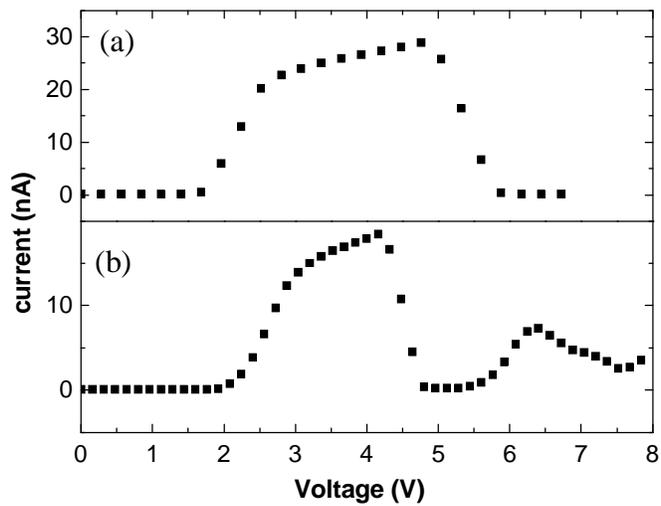

**Fig. 4**

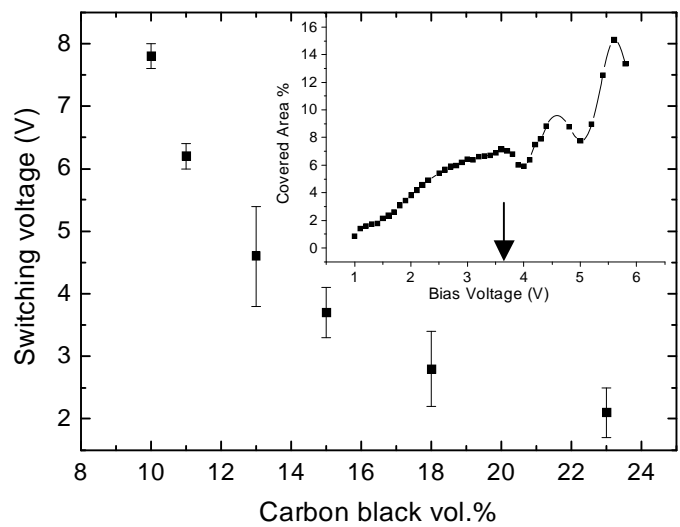

**Fig. 5**